\documentclass[superscriptaddress,amsmath,amssymb,aps,prl,twocolumn]{revtex4-1}

\usepackage{graphicx}
\usepackage{dcolumn}

\newcommand{\ket}[1]{{\left| {#1} \right\rangle}}

\newcommand{\HOMO}{\tilde{\mathrm{X}}}
\newcommand{\LUMO}{\tilde{\mathrm{A}}}
\newcommand{\SLUMO}{\tilde{\mathrm{B}}}
\newcommand{\LUMOtwo}{\tilde{\mathrm{C}}}

\begin{document}

\title{Franck-Condon tuning of optical cycling centers by organic functionalization}

\author{Claire E. Dickerson}\thanks{C.D. and H.G. contributed equally to this work.}
\affiliation{Department of Chemistry and Biochemistry, University of California Los Angeles}
\author{Han Guo}\thanks{C.D. and H.G. contributed equally to this work.}
\affiliation{Department of Chemistry and Biochemistry, University of California Los Angeles}
\author{Ashley J. Shin}
\affiliation{Department of Chemistry and Biochemistry, University of California Los Angeles}
\author{Benjamin L. Augenbraun}
\affiliation{Department of Physics, Harvard University}
\author{Justin R. Caram} \thanks{Corresponding author}
\affiliation{Department of Chemistry and Biochemistry, University of California Los Angeles}
\author{Wesley C. Campbell}\thanks{Corresponding author}
\affiliation{Department of Physics and Astronomy, University of California Los Angeles}
\author{Anastassia N. Alexandrova}\thanks{Corresponding author}
\affiliation{Department of Chemistry and Biochemistry, University of California Los Angeles}
\date{\today}

\begin{abstract}
Laser induced electronic excitations that spontaneously emit photons and decay directly to the initial ground state (``optical cycling transitions'') are used in quantum information and precision measurement for state initialization and readout.  To extend this primarily atomic technique to organic compounds, we theoretically investigate optical cycling of alkaline earth phenoxides and their functionalized derivatives. We find that optical cycle leakage due to wavefunction mismatch is low in these species, and can be further suppressed by using chemical substitution to boost the electron withdrawing strength of the aromatic molecular ligand through resonance and induction effects.
This provides a straightforward way to use chemical functional groups to construct optical cycling moieties for laser cooling, state preparation, and quantum measurement.

\end{abstract}

\maketitle
The use of isolated, complex systems in pure quantum states for computation, measurement, and sensing relies on the ability to determine the quantum state of the system.
This applies not only to state measurement, but also state preparation (and cooling), where initialization of the system to a pure quantum state is necessary to achieve quantum advantage \cite{DiVincenzoCriteria}. 

For state preparation and measurement (SPAM), spontaneously emitted photons following optical excitation are often employed as carriers of information (entropy) since they can be transported efficiently between systems that differ widely in temperature, mass, and size -- characteristics that comprise the gap between the isolated quantum system and its effectively classical environment.
However, the finite probability for detecting these photons (whose emission direction is usually randomized) means the cycle of excitation followed by spontaneous emission must be repeated many times (termed \textit{optical cycling}) to achieve single shot quantum state readout of single emitters.
Gas phase atoms driven by narrow-band lasers can facilitate this process through selection rules governing how their quantum numbers change during spontaneous emission, and have for many years been used in laser cooling, trapping, and SPAM \cite{Wineland1978RadiationPressure,Raab1987Trapping,Happer1972Optical,wineland1998experimental}.

Molecules, on the other hand, have internal vibrational degrees of freedom that need not be constrained by angular momentum selection rules and therefore can decay to  vibrationally excited levels of the ground state that lie below the excited state. 
This vibrational branching has largely precluded laser cooling of molecules and their use in quantum information, despite their highly desirable features \cite{DeMille2002Quantum,Yellin2006Schemes,Herrera2014InfraredDressed,Mallikarjun2016Prospects,Blackmore2019Ultracold,Ni2018Dipolar,Hudson2018Dipolar,Yu2019Scalable,Campbell2020DipolePhonon}.
For precision measurement, the statistical sensitivity of molecule-based approaches (such as the ACME $e$EDM search \cite{Andreev2018Improved,Panda2019Attaining}) is limited by the fact that, due to vibrational branching to dark states, only a small fraction of the molecules in the experiment are detected during readout.

However, recently, a few molecules have been experimentally shown to have sufficiently closed optical cycling transitions to allow laser cooling \cite{Shuman2009DiatomicRadiativeForce,Hummon20132D, Truppe2017Molecules,Kozyryev2017Sisyphus,Augenbaum2020LaserCooled,Ivanov2020Toward}. These molecules are characterized by vibrational branching ratios that strongly favor decay to a small number of ground-state vibrational levels, meaning only a handful of lasers are required to achieve optical cycling. Almost all of these molecules consist of an alkaline earth metal atom (M) ionically bonded to a molecular fragment in such a way that it optically behaves as a gas-phase M$^+$ cation radical.  Calculations have revealed that complex M-O-R (\textit{i.e.}~alkaline earth alkoxide) structures can be realized while retaining the ability to optically cycle \cite{Kozyryev2016Proposal,Ivanov2020Two,Klos2020Prospects, Augenbraun2020Molecular}.  However, the principles governing which ligands (R) will retain or even potentially promote optical cycling are not well understood, and searches for acceptable species currently rely heavily on trial and error with state of the art calculations for each candidate.

Building upon the M-O-R motif \cite{Klos2020Prospects,Ivanov2020Toward}, here we investigate functionalized phenyl rings for R and introduce a guiding principle by which the vibrational wavefunction overlap can be enhanced by straightforward chemical substitution within the molecular ligand. Using multireference wave functions and ground state and time-dependent density functional theory (TD-DFT) calculations \footnote{See Supplemental Material at [URL will be inserted by publisher] for details of the calculations.}, we investigate the Franck-Condon factors (FCFs, $q_{v^\prime,v^{\prime\prime}} \equiv |\langle v^\prime \ket{v^{\prime\prime}}|^2$, which in many cases approximate the vibrational branching ratio) of alkaline earth phenoxides. We show that (i) electronic transitions in Ca and Sr phenoxides are promising for optical cycling (see also \cite{Ivanov2020Toward}) and (ii) electron-withdrawing substituents on the phenyl ring make the M-O bond more ionic via induction and resonance effects. This substitution suppresses the FCF-induced vibrational branching of spontaneous emission roughly in proportion to the total electron withdrawing strength of the substituents.  In particular, making three $\mathrm{H}\!\rightarrow\mathrm{CF}_3$ substitutions on the ring of calcium phenoxide, despite nearly doubling the number of atoms in the molecule, boosts the FCF limit on the expected number of spontaneously emitted photons from 22 to more than 500, a level relevant for laser cooling~\cite{Kozyryev2017Sisyphus}. This technique should be applicable to a wide variety of molecules where the ionic character of the {M-O} bond in {MOR} can be manipulated from a distance via electron withdrawing organic functional groups in {R}.

\begin{figure}
    \centering
    \includegraphics[width=0.5\textwidth]{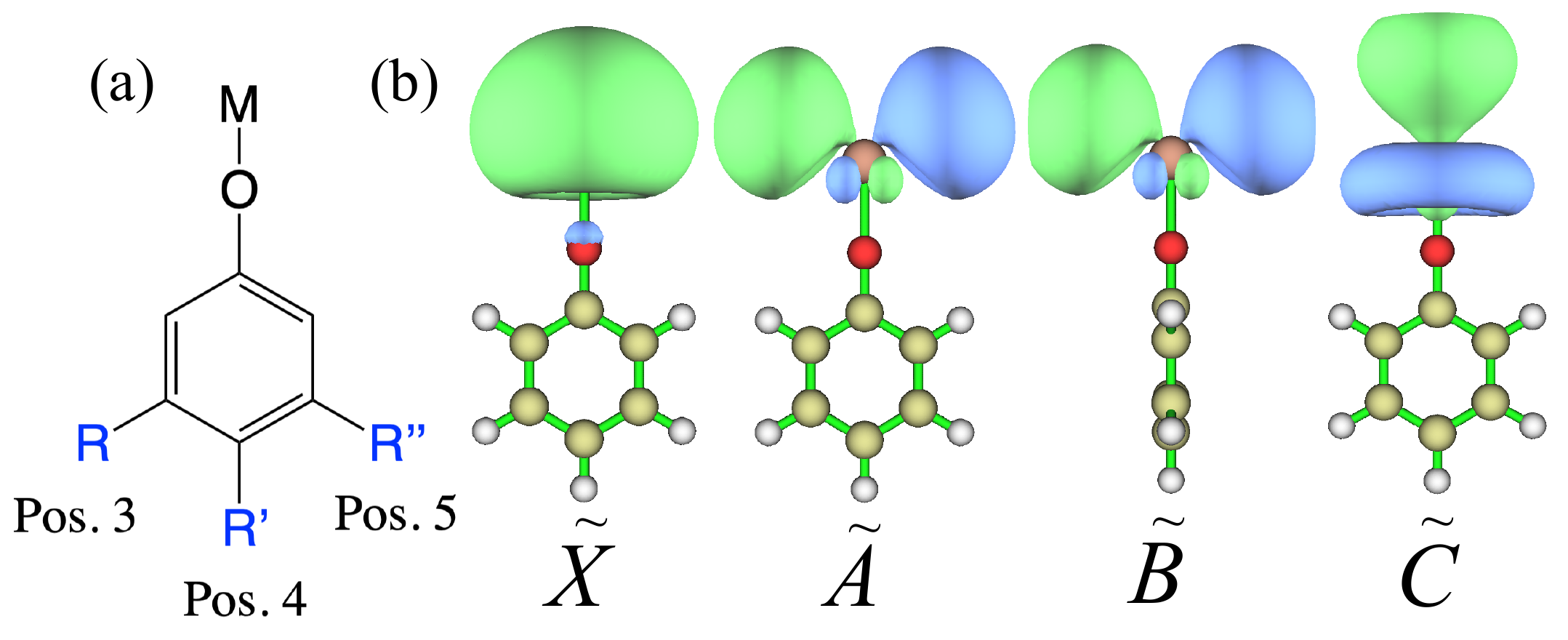}
    \caption{(a) Substitution positions investigated on the metal (Sr, Ca)-oxygen phenyl ring.   (b) Global minimum structures and molecular orbitals (isosurface value of 0.03) for SrOC$_{6}$H$_{5}$ show the atom-like character of the electron distribution for the ground and first three excited states.  The lack of electron density on the ligand suggests very little structural change will be involved in the electronic transitions.}
    \label{fig:MOs}
\end{figure}

We first describe our computational techniques and then show that the first three electronic transitions in Ca- and Sr-phenoxide have strong overlap between the ground and excited state vibrational wavefunctions.  We then show how the vibrational branching can be tuned by chemical substitution on the meta (3 and 5) and para (4) positions of the phenyl ring (see Fig.~\ref{fig:MOs}). The ability to control the Franck-Condon factors of large molecules (and, in particular, those containing benzene) may open the door to new applications in ultracold chemistry \cite{Krems2008Cold,Balakrishnan2016Perspective}, quantum information \cite{Yu2019Scalable,Carr2009Cold}, and precision measurement \cite{Kozyryev2017Precision,Jansen2014Perspective,Augenbaum2020LaserCooled}.

Many previous theoretical studies of optical cycling in molecules have used complete active space self-consistent field (CASSCF) and multireference configuration interaction (MRCI) methods in order to produce highly accurate results \cite{Hao2019High,Kang2016Suitability,Tohme2015Theoretical,Nayak2006Ab,Kozlov1997Enhancement}.  However, these methods generally become prohibitively expensive when applied to relatively large molecules.
DFT and TD-DFT, on the other hand, can compuationally  assess large species, but the accuracy of these methods for calculating vibrational branching is not well established.
Hence, we first benchmarked DFT and TD-DFT \cite{Furche2002Adiabatic,Liu2011Analyticala,Liu2011Analyticalb,Bauernschmitt1996Treatment} against both CASMRCI calculations and experimental measurements for the smallest MOR molecules, finding good agreement for the PBE0 hybrid functional \cite{Perdew1996Rationale} with the D3 dispersion corrections \cite{D3-Grimme2010}, def2-tzvppd basis set \cite{Rappoport2010def2-tzvppd} and the double harmonic approximation for Franck-Condon factors \cite{ezspectrum}.  While the accuracy of these methods for the large species considered below will remain an open question until they are tested by experiments, the FCFs we obtain from DFT and TD-DFT for the comparatively smaller alkaline earth hydroxides (MOH) and metholixdes (MOCH$_3$) are within 2\% of the experimentally measured branching ratios (see \cite{Note1} for details).

Using the techniques that produced the most accurate results for the smaller species, we first investigate the optical cycling properties of Ca- and Sr-phenoxide (\textit{i.e.}~a benzene molecule functionalized with an MO optical cycling center).  Figure \ref{fig:MOs} shows electron iso-surfaces for the highest occupied molecular orbial (HOMO, analogous to the ground state wavefunction of the unpaired valence electron on M) and the first few lowest unoccupied molecular orbitals (LUMOs, the same for the excited states).  In all cases shown, the electron density remains far from the molecular ligand, indicating that this valence electron plays very little role in the molecular bonds, a desirable property for suppressing vibrational branching.  In further support of the promise of these species for optical cycling, the orbitals themselves qualitatively resemble hydridized versions of the $s$ and $p_x$, $p_y$ and $p_z$ Cartesian-basis orbitals that constitute the optical cycling transition of the gas-phase atomic M$^+$ ion.  Transitions between the HOMO and the LUMO and LUMO+1 correspond roughly to the X$\,{}^2\Sigma^+ \! \leftrightarrow \! \mathrm{A} \, {}^2\Pi_{|\Omega|}$ fine structure doublet in the smaller, linear MOR species while the LUMO+2 is analogous to the B$ \, {}^2\Sigma^+$ state of those species. We label the electronic states as $\HOMO$, $\LUMO$, $\SLUMO$, and $\LUMOtwo$, in order of ascending energy.

To examine vibrational leakage from the optical cycle, we calculate the Franck-Condon factors (FCFs) for transitions from the vibrational ground state of the $\LUMO$ and $\LUMOtwo$ electronic excited states to the electronic ground state in the Born-Oppenheimer approximation.  Due to the lack of spectroscopic data and the difficulty of performing highly accurate calculations with large species (and since we will be interested in the \emph{marginal} effect of the the chemical substitutions, discussed below), we use the calculated FCF as a proxy for the true spontaneous emission branching.   Since these effects are expected to decrease the branching probability to the absolute ground state, we refer to $\eta_{0,0} \equiv \frac{q_{0,0}}{1-q_{0,0}}$ as the \textit{Franck-Condon limit} of the expected number of spontaneously emitted photons before a leakage event when no vibrational repumping lasers are applied.

For the $\LUMO \! \rightarrow \! \HOMO$ transitions in both Ca- and Sr-phenoxide, we find that the FCFs are indeed highly diagonal, as expected, with CaOC$_6$H$_5$ capable of emitting an average of $\eta_{0,0}\!=\!22$ photons before FCF-induced vibrational branching (we refer to the FCFs as ``diagonal'' if the Franck-Condon matrix $\mathbf{q}$ is approximately  equal to the identity). We also find that the $\LUMOtwo \! \rightarrow \HOMO$ transitions have even higher overlap, corresponding to both Ca- and Sr-phenoxide yielding $\eta_{0,0} > 150$ photons before FCF-induced vibrational branching. However, we find that vibronic coupling among the excited electronic states is likely to lead to perturbations that increase the vibrational branching ratios for $\SLUMO$ and $\LUMOtwo$ from those predicted by the unperturbed state analysis (see, e.g. \cite{mengesha2020branching, Baum2020establishing} and \cite{Note1}), and we therefore focus on the FCF-boosting effect of chemical substitution on $\LUMO \! \rightarrow \HOMO$, which is likely to be the most closed transition.

Fundamentally, the large values of $\eta_{0,0}$ attained by these species can be traced to the highly ionic nature of the M-O bond; the bonding electron of neutral M is pulled sufficiently far from the M$^+$ ion core that excitations of the remaining electron on the core do not perturb the bond.  This suggests (see also \cite{Ivanov2020Toward}) that \emph{if the electron withdrawing strength of the ligand can be increased, the FCF limit on the number of emitted photons would likewise increase  \cite{Ivanov2019Rational}}.  However, if electron withdrawing chemical groups are located too close to the metal atom, they pull on it and bend the bond, significantly degrading the diagonal FCFs. We therefore require an approach that allows the placement of electron-withdrawing chemical groups far from the M atom while still retaining sufficient chemical intercourse with M to increase the ionicity of the M-O bond.

\begin{figure}
    \centering
    \includegraphics[width=0.5\textwidth]{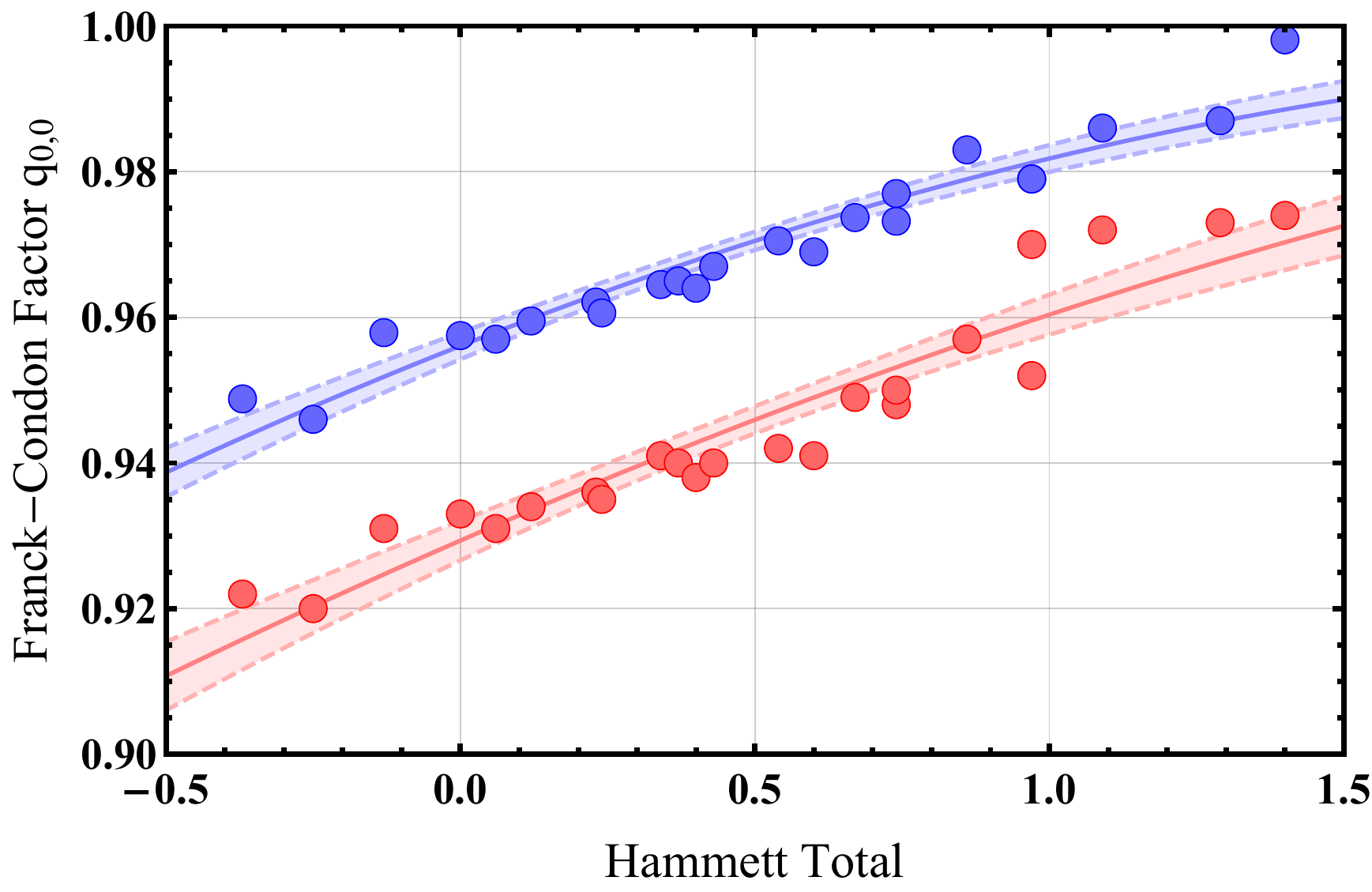}
    \caption{The calculated vibrationless $\LUMO \! \rightarrow \! \HOMO$ Franck-Condon factors for substituted MOC$_{6}$H$_{5}$ derivatives show a strong correlation with the total of the Hammett $\sigma$ constants of their substituents.  CaOR (SrOR) species are shown in blue (red).  Solid curves are fits to Gaussians centered at the $x$-intercept of the bond length change vs.~Hammett total trends (Table~\ref{table:FCF}).}
    \label{fig:FCFs}
\end{figure}

\begin{table}
\begin{ruledtabular}
%\begin{tabular*}{0.9 \columnwidth}{@{\extracolsep{\fill} } l | c  c  c}
\begin{tabular}{l | d  c c | c c}
Substituent& \multicolumn{1}{c}{\text{Hammett}} & \multicolumn{2}{c}{\text{CaOR} $\LUMO\! \rightarrow \! \HOMO$} & \multicolumn{2}{c}{\text{SrOR} $\LUMO\! \rightarrow \! \HOMO$}\\
for H in &\multicolumn{1}{c}{\mbox{Total}} & Ca-O &  FCF & Sr-O &  FCF\\
MOC$_6$H$_5$ & \multicolumn{1}{c}{$\sum \sigma$} &change ({\AA})& $q_{0,0}$&change ({\AA}) & $q_{0,0}$ \\
  \hline
4-OH  & -0.37 & 0.01809 & 0.949 & 0.02077 &  0.922\\
3,4-OH & -0.25 & 0.01756 & 0.946 & 0.02045 & 0.920 \\
3,4,5-OH & -0.13 & 0.01606 & 0.958 & 0.01938 & 0.931 \\
(none)  & 0 & 0.01680 & 0.958 & 0.01996 & 0.933 \\
4-F & 0.06 & 0.01657 & 0.957 & 0.01961 & 0.931 \\
3-OH & 0.12 & 0.01622 & 0.960 & 0.01956 & 0.934 \\
4-Cl  & 0.23 & 0.01552 & 0.962 & 0.01896 & 0.936 \\
3,5-OH & 0.24 & 0.01568 & 0.961 & 0.01919 & 0.935 \\
3-F  & 0.34 & 0.01497& 0.965 & 0.01855 & 0.941 \\
3-Cl  & 0.37 & 0.01461 & 0.965 & 0.01832 & 0.940 \\
3,4-F & 0.40 & 0.01479 & 0.964 & 0.01826 & 0.938\\
3-CF$_{3}$ & 0.43 & 0.01376 & 0.967 & 0.01765 & 0.940\\
4-CF$_{3}$  & 0.54 & 0.01358 & 0.971 & 0.01767 & 0.942 \\
3,4-Cl & 0.60 & 0.01357 & 0.969 & 0.01754 & 0.941 \\
3,5-F  & 0.67 & 0.01302 & 0.974 & 0.01707 & 0.949\\
3,4,5-F  & 0.74 & 0.01290 & 0.973 & 0.01686 & 0.948 \\
3,5-Cl  & 0.74 & 0.01269 & 0.977 & 0.01662 & 0.950\\
3,5-CF$_{3}$ & 0.86 & 0.01022 & 0.983 & 0.01498 & 0.957\\
3,4,5-Cl  & 0.97 & 0.01159 & 0.979 & 0.01610 &  0.952\\
3,4-CF$_{3}$ & 0.97 & 0.01009 & 0.979 & 0.01296 & 0.970 \\
3,5-CF$_{3}$-4-Cl & 1.09 & 0.00927 & 0.986 & 0.01236 & 0.972 \\
3,5-Cl-4-CF$_{3}$ & 1.29 & 0.00882 & 0.987 & 0.01231 & 0.973\\
3,4,5-CF$_3$  & 1.40 & 0.00290 & 0.998 & 0.01198 & 0.974\\
\end{tabular}
\end{ruledtabular}
\caption{\label{table:FCF}Calculated changes in the M-O bond length (positive indicates bond lengthening upon emission) and Franck-Condon factors for the $\LUMO \! \rightarrow \! \HOMO$ transitions in M-phenoxide with various functional groups on the 3, 4, and 5 positions of the phenyl ring.  The Hammett $\sigma$ constants of each substituent are summed to indicate the additional electron withdrawing strength contributed by the substitution.}
\end{table}

For this, we employ substituents at the 3, 4, and 5 positions of the phenyl ring which withdraw electrons via resonance and inductive effects and influence the M-O bond without compromising its linearity.
As a predictor of their expected influence on the M-O bond ionicity, we apply the concept of Hammett $\sigma$ constants \cite{Hammett1937}, dimensionless parameters that are empirically determined from ionization of organic acids in liquid and have been tabulated for many functional groups and substitution locations (see e.g.~\cite{Hansch1991Survey}).  Despite the seeming conceptual disconnect between the chemistry of species in solution and optical cycling, we show that the Hammet $\sigma$ constants effectively provide a guide for the effect of substituents on Franck-Condon overlap since they quantify the electron donating or withdrawing effect of each substitution. Roughly speaking, positive Hammett constants indicate electron withdrawing strength with negative constants indicating electron donation, so we therefore expect large, positive totals for the Hammett constants of the substituted functional groups to suppress FCF-induced vibrational branching.

Figure~\ref{fig:FCFs} and Table~\ref{table:FCF} show calculated vibrationless (i.e.~from absolute vibrational ground state to absolute vibrational ground state) Franck Condon factors $q_{0,0}$ on $\LUMO \! \rightarrow \! \HOMO$ as a function of the total of the Hammett constants for a variety of functional groups added to the phenyl rings of Ca- and Sr-phenoxide. For this we chose to examine OH, Cl, F, and CF$_3$ in all possible configurations of the 3,4, and 5 positions, as well as several mixtures of these. The FCF-limited optical cycle closure shows a clear positive correlation with Hammett constant total as various substitutions are made that impact the ionic nature of the M-O bond, in accordance with the principle described above.  In particular, the substitution of three CF$_3$ groups for three hydrogens on the far side of the ring in CaOC$_6$H$_5$ increases $q_{0,0}$ from 0.958 to 0.998, a boost in the FCF-limit for the expected number of photons ($\eta_{0,0}$) by more than a factor of $20\!\,\times$ compared to the unaltered variant.

The effect of chemical substitution on vibrational branching in these species can be traced largely to their geometry.  Table~\ref{table:FCF} shows that the length change of the M-O bond for the $\LUMO \! \rightarrow \! \HOMO$ transition is approximately linearly correlated with the Hammett total of the substituents.   In all cases, the largest geometry change from ground to excited state was the M-O bond length.  All other bond lengths remained unchanged from ground to excited state within the calculated accuracy (0.005 angstroms).  Extrapolation of the linear trend to zero bond length change can be used to build a simplified model for how the FCFs should depend upon the Hammett total.  Since vibrational ground states are approximately Gaussian and the transition's bond length change is linear in the Hammett total, the vibrationless FCFs ($q_{0,0}$) will be Gaussian in Hammett total.  The solid curves in Fig.~\ref{fig:FCFs} are Gaussian fits to the calculated points, which appear consistent with this model.

\begin{figure}
    \centering
    \includegraphics[width=0.5\textwidth]{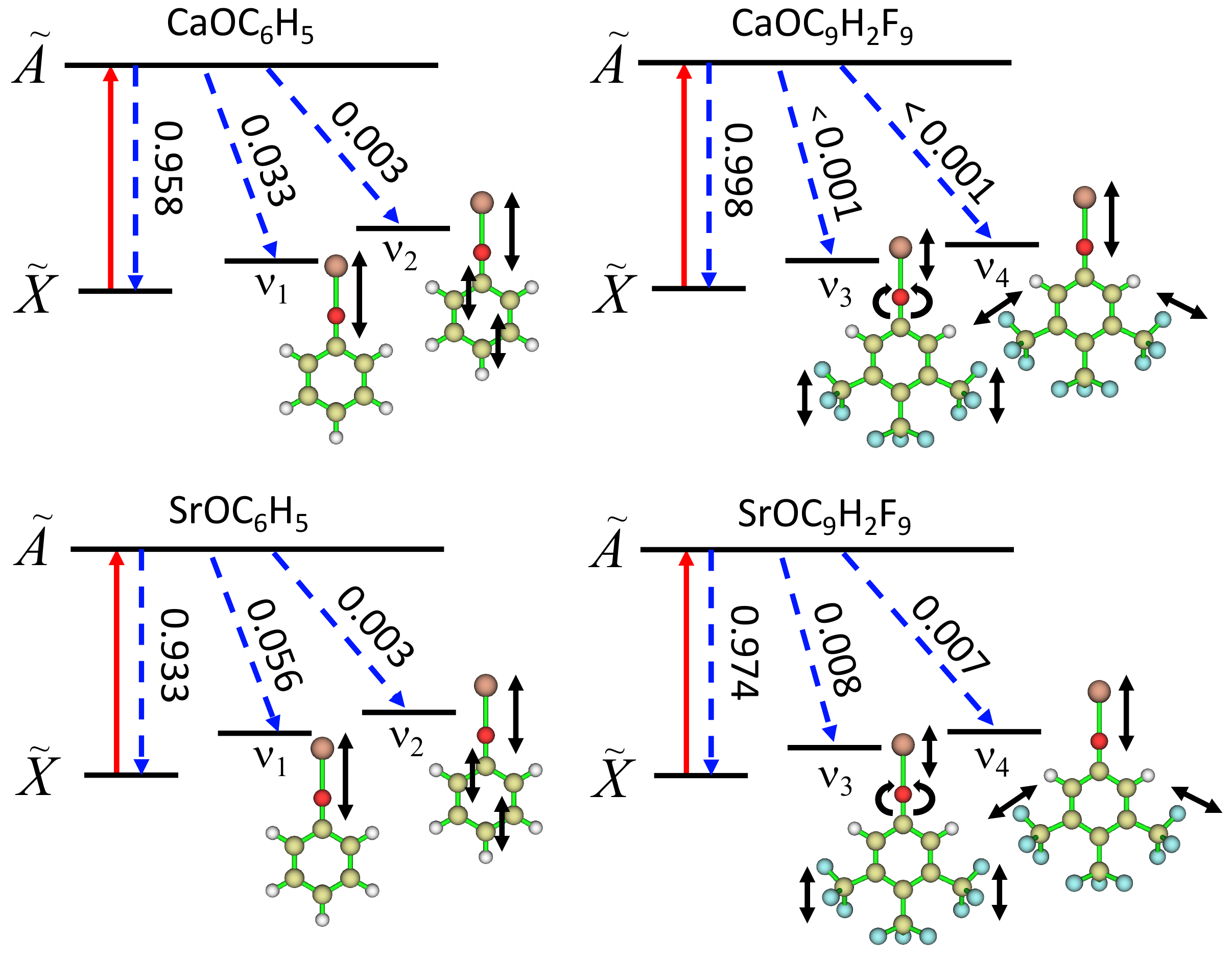}
    \caption{Photon cycling scheme with an excitation (red) to the first excited electronic state ($\LUMO$), and decay (blue) to the ground electronic state ($\HOMO$).  The FCFs are shown along with each decay.}
    \label{fig:modes}
\end{figure}

In all MOC$_{6}$H$_{5}$ derivatives, the $\LUMO \! \rightarrow \! \HOMO$ transition's off-diagonal FCFs were dominated by a normal mode strongly associated with stretching of {M-O.}  Figure \ref{fig:modes} shows the diagonal FCF  (the fundamental transition) and the largest two off-diagonal FCFs, labeled with their associated normal modes for the unsubstituted and trifluoromethyl-substituted SrOC$_{6}$H$_{5}$ and CaOC$_{6}$H$_{5}$.
For both SrOC$_{6}$H$_{5}$ and  CaOC$_{6}$H$_{5}$, the largest leakage pathways are normal modes with almost entirely M-O stretch character.  As more electron-withdrawing susbstituents are added, this isolated stretch mode incorporates more and more bending behavior, until the largest electron-withdrawing group case, MOC$_{9}$H$_{2}$F$_{9}$, has a leakage pathway dominated by a vibrational mode with combined M-O stretching and bending character.  In addition, analysis of second-order coupling to nearby vibronic levels predicts induced loss channels smaller than $10^{-3}$ on $\LUMO \! \rightarrow \! \HOMO$, suggesting that these FCFs can be used as a guide to investigate optical cycling in these species since they are all less than $\approx 0.999$ \cite{Note1}.

The use of electron-withdrawing functional groups to boost the FCF of the optical cycling transitions in large molecules relies on two basic properties of the metal and ligand.  First, the (possibly substituted) ligand's HOMO/LUMO gap must be large enough to fit the electronic excitation of the metal in the gap.   For example, benzene and adamantane have a naturally large HOMO/LUMO gap which can easily append a metal with an unpaired electron such as Sr or Ca, and can be decorated with electron-withdrawing substituents.  The new HOMO/LUMO gap for these M-O-R species becomes the metal to metal electronic transition, creating an isolated electronic transition.
Second, it is important that electron withdrawing substituents do not delocalize the optically active electron.
For example, we find that if 4-NO$_{2}$ is substituted on the phenyl group, it promotes delocalization through the $\pi$ system of its molecular orbitals, unlike trifluoromethyl substituents, and spreads the electronic wavefunction across NO$_{2}$ and the benzene ring. 
This reorders the unoccupied orbitals such that the new LUMO is the electron density delocalized on the benzene ligand instead of localized on the metal.  This can also be seen as electron density mixing of metal and NO$_{2}$ in natural transition orbitals \cite{Note1}.
As a result, substituents that favor delocalized $\pi$ bonds are poor candidates for FCF tuning.

The technique presented here of using chemical substitution to bolster optical cycling introduces a principle for informed design of species for quantum information and precision measurement applications.  By using Hammett constants to choose electron-withdrawing substituents, the expectation that increasing the number of vibrational modes (and therefore decay channels) will compromise optical cycle closure can be circumvented.  Indeed, we have shown here that optical cycling can actually be improved by adding more complexity to the organic ligand by deliberately utilizing its size and aromatic properties to allow these functional groups to operate far from the optical cycling center --- a capability that only large molecules can provide. 
While we have focused here on a few particular features of large molecules that can promote optical cycling, the observation that new features can emerge as complexity increases supports the claim that more aspects of polyatomic molecules are likely be identified in the future to allow increased quantum control of molecular species.

\textit{Acknowledgments }This work was supported by the U.S. Department of Energy, Office of Science, Basic Energy Sciences, under Award \#DE-SC0019245.  The authors acknowledge helpful discussions with Eric Hudson.

\bibliography{FranckCondonTuning}

\end{document}